\documentclass[%
 preprint,longbibliography,
 amsmath,amssymb,
 aps,prd
]{revtex4-1}

\newcommand\fft[2]{\frac{#1}{#2}}
\newcommand\ft[2]{{\textstyle\frac{#1}{#2}}}
\newcommand\ffd[2]{{\displaystyle\frac{#1}{#2}}}
\newcommand\nn{\nonumber}

\begin{document}

\preprint{LCTP-18-10}

\title{The Weyl Anomaly from the 6D Superconformal Index}

\author{James T. Liu}
\email{jimliu@umich.edu}
\affiliation{Leinweber Center for Theoretical Physics, Randall Laboratory of Physics, The University of Michigan, Ann Arbor, MI 48109-1040}

\author{Brian McPeak}
\email{bmcpeak@umich.edu}
\affiliation{Leinweber Center for Theoretical Physics, Randall Laboratory of Physics, The University of Michigan, Ann Arbor, MI 48109-1040}

\begin{abstract}

We explore the connection between the holographic Weyl anomaly and the superconformal index in six-dimensional $(1,0)$ theories. Using earlier results from holographic computations of the $\mathcal O(1)$ contributions $\delta a$ and $\delta(c-a)$ to the corresponding six-dimensional Weyl anomaly coefficients, we derive a pair of differential operators that extracts these values from the large-$N$ single-trace index.  In doing so, we also highlight the structure of these corrections in terms of group theory invariants of the superconformal representations.

\end{abstract}

\maketitle


\section{Introduction}

Superconformal field theories have a rich structure that often make them amenable to study, even at strong coupling.  Moreover, the combination of localization and holography has led to a much deeper understanding of such theories in the large-$N$ limit.  An example of this is the connection between the superconformal index and central charges of the theory.  The superconformal index is a refined Witten index \cite{Romelsberger:2005eg,Kinney:2005ej} that encodes information on the shortened spectrum of the theory.  While it depends on the details of the spectrum, its ``high-temperature'' limit exhibits universal behavior that is governed by the central charges of the superconformal theory \cite{DiPietro:2014bca,Ardehali:2015hya,Ardehali:2015bla,DiPietro:2016ond}.  This behavior was explicitly demonstrated in AdS$_5$/CFT$_4$ holography, where the $\mathcal O(1)$ contribution to the $a$ and $c$ central charges were obtained by acting with appropriate differential operators on the large-$N$ single-trace index \cite{Ardehali:2014zba,Ardehali:2014esa}.

The observation of \cite{Ardehali:2014zba,Ardehali:2014esa} is that while the leading order contribution to the central charges depends on the tree-level supergravity action, the $\mathcal O(1)$ corrections, $\delta a$ and $\delta c$, arise from a one-loop computation.  In principle, this is obtained as a sum over the full spectrum of states in the holographic dual.  However, explicit calculations in AdS$_5$ demonstrate that the contribution from states in long representations cancel completely, leaving only shortened representations contributing to $\delta a$ and $\delta c$ \cite{Beccaria:2014xda}.  It is for this reason that the holographic central charges can be extracted from the index.

In this paper, we extend the relation of holographic central charges to the superconformal index in the case of AdS$_7$/CFT$_6$.  Six-dimensional superconformal field theories are noteworthy as six is the highest possible dimension for superconformal invariance, and such theories can be reduced on Riemann surfaces to give a large class of theories in four dimensions.  Of course, less is known about six-dimensional superconformal field theories, and moreover the present situation is complicated by the fact that we have to consider four central charges, $\{a, c_1, c_2, c_3\}$.  Assuming $(1,0)$ supersymmetry, these charges satisfy the relation $c_1-2c_2+6c_3=0$, so there are only three independent coefficients, which we denote $a$, $c$ and $c'$ according to \cite{Liu:2017ruz}
\begin{equation}
    c_1=96(c+c'),\qquad c_2=24(c-c'),\qquad c_3=-8(c+3c').
\end{equation}
The $\mathcal O(1)$ contribution to $a$ was obtained in \cite{Beccaria:2014qea} by computing the heat kernel for arbitrary spin fields on global AdS$_7$ with $S^6$ boundary.  Unfortunately, this computation does not provide any information on $c$ or $c'$, since $S^6$ is conformally flat.  

In order to go beyond just $\delta a$, the holographic computation has to be extended away from the case of a conformally flat boundary.  This was done in \cite{Liu:2017ruz} for a Ricci-flat boundary using the functional Schr\"odinger method \cite{Mansfield:1999kk,Mansfield:2000zw,Mansfield:2002pa,Mansfield:2003gs,Mansfield:2003bg}.  This computation gives the $\mathcal O(1)$ contribution to $c-a$ for states with spins up to two, but no information on $c'$ as the corresponding Weyl invariant vanishes up to a total derivative on Ricci-flat backgrounds.  Of course, one would ideally want a general expression for both $c$ and $c'$ and for arbitrary spin states.  However, use of the functional Schr\"odinger method to compute the one-loop holographic Weyl anomaly appears to be restricted to maximum spin two on Ricci-flat backgrounds.  The failure of this method on general backgrounds is related to the failure of higher-derivative kinetic operators to factor in non-Ricci flat manifolds \cite{Nutma:2014pua,Grigoriev:2016bzl}, and the issue of higher spins appears to be related to the presence of new mixed kinetic terms which only show up for higher spin multiplets \cite{Grigoriev:2016bzl,Beccaria:2017nco}.

Given our knowledge of $\delta a$ and $\delta(c-a)$ for $(1,0)$ theories, we demonstrate below how they may be obtained from the large-$N$ single-trace index.  In particular, we construct differential operators that extract $\delta a$ and $\delta(c-a)$ from the index in the high-temperature limit.  The expression for $\delta a$ is fully constrained, while that for $\delta(c-a)$ has one undetermined coefficient related to our lack of knowledge of the $\mathcal O(1)$ holographic Weyl anomaly beyond spin two.

\section{The Superconformal Index for the $(1,0)$ Theory}

The four-dimensional superconformal index was introduced in \cite{Romelsberger:2005eg,Kinney:2005ej} and generalized to additional dimensions in \cite{Bhattacharya:2008zy}.  Before discussing the index, we first briefly review the $\mathcal N=(1,0)$ theory.  Six dimensions is the highest dimension that admits superconformal symmetry, and $(1,0)$ supersymmetry is minimal.  The superconformal algebra decomposes as $OSp(8^*|2)\supset SO(2,6)\times SU(2)_R\supset U(1)_\Delta\times SU(4)\times SU(2)_R$.  Unitary representations may be labeled by conformal dimension $\Delta$, $SU(4)$ Dynkin labels $(j_1,j_2,j_3)$ and the $SU(2)_R$ label $k$ (with `spin' $k/2$).

Long representations of $(1,0)$ have $\Delta>\fft12(j_1+2j_2+3j_3)+2k+6$, while short representations fall into four categories, comprising one regular and three isolated short multiplets.  The shortening conditions are given by \cite{Bhattacharya:2008zy,Buican:2016hpb,Cordova:2016emh}
\begin{align}
    A[j_1,j_2,j_3;k]:&&&\Delta=\ft12(j_1+2j_2+3j_3)+2k+6,\nn\\
    B[j_1,j_2,0;k]:&&&\Delta=\ft12(j_1+2j_2)+2k+4,\nn\\
    C[j_1,0,0;k]:&&&\Delta=\ft12j_1+2k+2,\nn\\
    D[0,0,0;k]:&&&\Delta=2k.
    \label{eq:10short}
\end{align}
Long representations are generated by the action of all 16 real supercharges and have states with dimensions ranging from $\Delta$ to $\Delta+4$, while the successive shortened representations generically have dimensions going up to $\Delta+7/2$, $\Delta+3$, $\Delta+5/2$ and $\Delta+2$, respectively.  The latter $D$ multiplets are generated by eight supercharges and are half-BPS.

We now turn to the six-dimensional $(1,0)$ index which was introduced in \cite{Bhattacharya:2008zy} as
\begin{align}
    \mathcal{I}(p,q,s) = \mathrm{Tr}_{\mathcal{H}}(-1)^{j_1 + j_3}e^{- \beta \delta} q^{\Delta - \frac{1}{2}k}s^{j_1}p^{j_2} ,
    \label{eq:index}
\end{align}
where $\delta = \Delta - 2k - \frac{1}{2}(j_1 + 2 j_2 + 3 j_3)$.  Recall here that $(j_1,j_2,j_3)$ labels the $SU(4)$ Lorentz representation, and that $j_1+j+3$ represents the fermion number.  In particular, the index is a Witten index refined by fugacities $q$, $s$ and $p$ associated with the charges $\Delta-k/2$, $j_1$ and $j_2$ that commute with the supercharge $\mathcal Q$ used to define the index.  While the trace is {\it a priori} over all states in the spectrum, only those satisfying $\delta=0$ will contribute.  Thus the index is actually independent of $\beta$, and only receives contributions from shortened multiplets.

Since we are motivated by the holographic dual, our main interest is on the single-trace index, which corresponds to the single particle spectrum.  In this case, the expression (\ref{eq:index}) has a particularly simple form.  To see this, we first note that the charges $j_1$ and $j_2$ in (\ref{eq:index}) are $SU(3)$ weights corresponding to the breaking of $SU(4)$ by the defining supercharge $\mathcal Q$.  As a result, the index can be decomposed as a sum over $SU(3)$ characters $\chi_{(j_1,j_2)}(s,p)$ given by the Weyl character formula:
\begin{equation}
\chi_{(j_1,j_2)}(s,p)= \frac{\begin{pmatrix}{s^{j_1+1}p^{j_2+1}-s^{-j_2-1}p^{-j_1-1}  
    + s^{j_2+1}p^{-j_1-j_2-2}-s^{j_1+j_2+2}p^{-j_2-1} }\\{
    +s^{-j_1-j_2-2}p^{j_1+1} 
    -s^{-j_1-1}p^{j_1+j_2+2}} \end{pmatrix}}{\left(\sqrt{sp }-\ffd{1}{\sqrt{ sp}}\right)\left(\ffd{s}{\sqrt{p}}-\ffd{\sqrt{p}}{s}\right)\left(\ffd{p}{\sqrt{s}}-\ffd{\sqrt{s}}{p}\right) 
   }
\end{equation}
Moreover, for a given representation, the index receives contributions from both superconformal primaries and their descendants.  The contributions from the latter are captured by the denominator factor
\begin{equation}
    \fft1{\mathcal D(p,q,s)}=\fft1{(1-qs^{-1})(1-qp)(1-qs/p)}=1+q\chi_{(0,1)}(s,p)+q^2\chi_{(0,2)}(s,p)+\cdots.
\end{equation}
As a result, the single-trace index for a given short representation takes the form
\begin{equation}
    \mathcal I(p,q,s)\sim q^{\Delta-\fft{k}2}\fft{\chi(s,p)}{\mathcal D(p,q,s)},
\end{equation}
for some appropriate $SU(3)$ character $\chi(s,p)$.  The indices were worked out on a representation by representation basis in \cite{Buican:2016hpb}, and we summarize the results in Table~\ref{tb:index}.

\begin{table}[t]
\begin{center}
  \begin{tabular}{| l | l | l |}
    \hline
    \ Multiplet & \ Shortening Condition & \ $\mathcal D(p,q,s)\mathcal{I}_R(p,q,s)$ \ \\
    \hline
    \hline
    \ $A[j_1,j_2,j_3;k]$ \ & \ $\Delta=\ft12(j_1+2j_2+3j_3)+2k+6$ \ & \ $(-1)^{j_1 + j_3+1} \, q^{\Delta-\fft12k} \, \chi_{(j_1, j_2)}(s,p)$ \ \\ \hline
    \ $B[j_1,j_2,0;k]$ \ & \ $\Delta=\ft12(j_1+2j_2)+2k+4$ & \ $(-1)^{j_1} \, q^{\Delta-\fft12k} \, \chi_{(j_1, j_2+1)}(s,p)$ \ \\ \hline
    \ $C[j_1,0,0;k]$ \ & \ $ \Delta=\ft12j_1+2k+2$ & \ $(-1)^{j_1+1} \, q^{\Delta-\fft12k} \, \chi_{(j_1+1, 0)}(s,p)$ \ \\ \hline
    \ $D[0,0,0;k]$ \ & \ $\Delta=2k$ & \ $ \,  q^{\Delta-\fft12k} \, \chi_{(0, 0)}(s,p)$ \ \\ \hline
  \end{tabular}
\end{center}
\caption{Contribution to the single-trace index for $(1,0)$ short multiplets with Dynkin labels $(j_1, j_2, j_3)$, conformal weight $\Delta$, and $R$-charge $k$.  Here we are taking generic values for $j_1,j_2,j_3$ and $k$; some special cases arise at small values of the quantum numbers.
\label{tb:index}}
\end{table}

\section{One-loop holographic central charges}

For $(1,0)$ theories with a large-$N$ dual, we generally expect the central charges to scale as $\mathcal O(N^3)$.  Holographically, the leading contribution comes from the tree-level bulk action \cite{Henningson:1998gx}. Sub-leading terms of $\mathcal O(N)$ arise from $\alpha'^3R^4$ corrections and terms of $\mathcal O(1)$ from the one-loop determinant.  It is the latter terms that we focus on.

\subsection{The $\mathcal O(1)$ shift $\delta a$}

We first examine the $\mathcal O(1)$ contribution $\delta a$ to the $a$ central charge.  This was evaluated in \cite{Beccaria:2014qea} for an arbitrary representation of the $SO(2,6)$ conformal group labeled by $D(\Delta,j_1,j_2,j_3)$ by computing the heat kernel group theoretically on global AdS$_7$. The result can be expressed as
\begin{align}
    \delta a(\Delta,j_1,j_2,j_3)&=\fft{(-1)^{j_1+j_3}(\Delta-3)}{2^5\cdot6!}\biggl[\fft1{21}(\Delta-3)^6d(j_1,j_2,j_3)\nn\\
    &\kern4em-(\Delta-3)^4\left(I_2(j_1,j_2,j_3)+\fft13d(j_1,j_2,j_3)\right)\nn\\
    &\kern4em+(\Delta-3)^2\biggl(\fft{70}{51}I_4(j_1,j_2,j_3)+\fft{75}{17}\fft{I_2(j_1,j_2,j_3)^2}{d(j_1,j_2,j_3)}+\fft{50}{17}I_2(j_1,j_2,j_3)\nn\\
    &\kern9em+\fft49d(j_1,j_2,j_3)\biggr)
    -\fft{75}4\fft{I_3(j_1,j_2,j_3)}{d(j_1,j_2,j_3)}\biggr],
\label{eq:btda}
\end{align}
where the sign factor $(-1)^F=(-1)^{j_1+j_3}$ distinguishes between bosons and fermions.  Here we have rewritten the expression of \cite{Beccaria:2014qea} in terms of $SU(4)$ invariants where
\begin{equation}
    d(j_1,j_2,j_3)=\fft1{12}(j_1+1)(j_2+1)(j_3+1)(j_1+j_2+2)(j_2+j_3+2)(j_1+j_2+j_3+3),
\end{equation}
is the dimension of the representation and the $I_a$'s are indices
\begin{align}
    I_2(j_1,j_2,j_3)&=\fft1{60}d(j_1,j_2,j_3)[3j_1^2+12j_1+4j_1j_2+2j_1j_3+4j_2^2+4j_2j_3+16j_2+3j_3^2+12j_3],\nn\\
    I_3(j_1,j_2,j_3)&=\fft1{60}d(j_1,j_2,j_3)(j_1-j_3)(j_1+j_3+2)(j_1+2j_2+j_3+4),\nn\\
    I_4(j_1,j_2,j_3)&=\fft1{420}d(j_1,j_2,j_3)[3 j_1^4 + 8 j_1^3 j_2 + 2 j_1^2 j_2^2 - 12 j_1 j_2^3 - 6 j_2^4 + 4 j_1^3 j_3 + 2 j_1^2 j_2 j_3\nn\\
    &\kern4em- 18 j_1 j_2^2 j_3 - 12 j_2^3 j_3 - 4 j_1^2 j_3^2 + 2 j_1 j_2 j_3^2 + 2 j_2^2 j_3^2 + 4 j_1 j_3^3 + 8 j_2 j_3^3 + 3 j_3^4\nn\\
    &\kern4em+24 j_1^3 + 30 j_1^2 j_2 - 50 j_1 j_2^2 - 48 j_2^3 + 6 j_1^2 j_3 - 28 j_1 j_2 j_3 - 50 j_2^2 j_3 + 6 j_1 j_3^2\nn\\
    &\kern4em+ 30 j_2 j_3^2 + 24 j_3^3
    +54 j_1^2 - 34 j_1 j_2 - 122 j_2^2 - 2 j_1 j_3 - 34 j_2 j_3 +  54 j_3^2\nn\\
    &\kern4em+24 j_1 - 104 j_2 + 24 j_3
 ],
\end{align}
normalize to unity for the fundamental $(1,0,0)$ representation.

For the $(1,0)$ superconformal case, we compute the shift $\delta a$ for each supermultiplet by summing (\ref{eq:btda}) over the individual states comprising the representation.  The multiplet structure has been worked out explicitly in \cite{Buican:2016hpb,Cordova:2016emh}, and using those results, we may obtain $\delta a$ for each type of shortened multiplet given in (\ref{eq:10short}):
\begin{equation}
    \delta a=\begin{cases}
    (-1)^{j_1+j_3+1}\mathcal A(j_1,j_2,\Delta-\ft12k),&A[j_1,j_2,j_3;k];\nn\\
    (-1)^{j_1}\mathcal A(j_1,j_2+1,\Delta-\ft12k),&B[j_1,j_2,0;k];\nn\\
    (-1)^{j_1+1}\mathcal A(j_1+1,0,\Delta-\ft12k),&C[j_1,0,0;k];\nn\\
   \mathcal A(0,0,\Delta-\ft12k),&D[0,0,0;k].
    \end{cases}
    \label{eq:hda}
\end{equation}
Here $\mathcal A(j_1,j_2,\hat\Delta)$ has the universal form
\begin{align}
    2^5\cdot6!\mathcal A(j_1,j_2,\hat\Delta)&=-10\left(\fft43\hat\Delta-2\right)^4 d(j_1,j_2)+20\left(\fft43\hat\Delta-2\right)^2[4I_2(j_1,j_2)+d(j_1,j_2)]\nn\\
    &\quad+\fft{530}9\left(\fft43\hat\Delta-2\right)I_3(j_1,j_2)-\fft{80}9[I_{2,2}(j_1,j_2)+3I_2(j_1,j_2)]-\fft{11}3d(j_1,j_2),
    \label{eq:calA}
\end{align}
where
\begin{equation}
    d(j_1,j_2)=\ft12(j_1+1)(j_2+1)(j_1+j_2+2),
\end{equation}
is the dimension of the $SU(3)$ representation and the $I_a$'s are indices
\begin{align}
    I_2(j_1,j_2)&=\fft1{12}d(j_1,j_2)[j_1^2+3j_1+j_1j_2+j_2^2+3j_2],\nn\\
    I_3(j_1,j_2)&=\fft1{60}d(j_1,j_2)(j_1-j_2)(j_1+2j_2+3)(2j_1+j_2+3),\nn\\
    I_{2,2}(j_1,j_2)&=\fft35I_2(j_1,j_2)\left(8\fft{I_2(j_1,j_2)}{d(j_1,j_2)}-1\right),
    \label{eq:SU3index}
\end{align}
normalized to unity for the fundamental $(1,0)$ representation.  Since $SU(3)$ has rank two, it only has two independent Casimir invariants, with corresponding indices $I_2$ and $I_3$.  Therefore the fourth order index $I_{2,2}$ is not independent, but can be decomposed in terms of $I_2$ as indicated above.

It is now apparent that the structure of the holographic $\delta a$ in (\ref{eq:hda}) closely resembles that of the single-trace index as shown in Table~\ref{tb:index}.  This connection can be made precise by associating the factor $q^{\Delta-\fft12k}\chi_{(j_1,j_2)}(s,p)$ in the index with the anomaly function $\mathcal A(j_1,j_2,\Delta-\fft12)$.  This is easily done once we realize that the indices can be obtained from the $SU(3)$ character $\chi_{(j_1,j_2)}(s,p)$.  The relation is not unique, but one possibility is to take
\begin{align}
    d(j_1,j_2)&=\chi_{(j_1,j_2)}(s,p)\big|_{s=p=1},\nn\\
    I_2(j_1,j_2)&=\ft{1}{2}
    (s \partial_s)^2 \chi_{(j_1, j_2)}(s,p)\big|_{s=p=1},\nn \\
    I_3(j_1,j_2)&= (p \partial_p)(s \partial_s)^2 \chi_{(j_1, j_2)}(s,p)\big|_{s=p=1},\nn\\
    I_{2,2}(j_1,j_2)&=\ft{1}{2}(s\partial_s)^4 \chi_{(j_1, j_2)}(s,p)\big|_{s=p=1}.
    \label{eq:Ichi}
\end{align}
The reason we have left $I_{2,2}$ in the $\delta a$ expression (\ref{eq:calA}) is now apparent, as it can be obtained directly from the character as opposed to the square of $I_2$.

Combining the above observations, we are now led to the final expression relating $\delta a$ to the single-trace index
\begin{align}
    \delta a&=\fft1{2^5\cdot6!}\biggl[-10\left(\fft43q\partial_q-2\right)^4+20\left(\fft43q\partial_q-2\right)^2(4\hat I_2+1)+\fft{530}9\left(\fft43q\partial_q-2\right)\hat I_3\nn\\
    &\kern4.3em-\fft{80}9(\hat I_{2,2}+3\hat I_2)-\fft{11}3\biggr]\mathcal D(p,q,s)\mathcal I(p,q,s)\bigg|_{p=q=s=1}.
    \label{eq:afromI}
\end{align}
Here the $\hat I_a$'s correspond to the differential operators used in (\ref{eq:Ichi}) to obtain the indices from the group character.

\subsection{The $\mathcal O(1)$ shift $\delta (c-a)$}

We now turn to consideration of holographic $\delta(c-a)$.  So far, this has only been worked out for maximum spin-two multiplets, so the information is necessarily incomplete.  Nevertheless, there is still a useful connection to be made, and the data is shown in Table~\ref{tb:c-aindex}.  Noting that the relevant $SU(3)$ representations are the singlet, triplet and anti-triplet, and that the indices, (\ref{eq:SU3index}), are normalized to unity for the triplet, we obtain the expression
\begin{equation}
    \delta(c-a)=\fft1{2^5\cdot6!}\left[-90\left(\fft43q\partial_q-2\right)\hat I_3+1+\lambda(\hat I_{2,2}-\hat I_2)\right]\mathcal D(p,q,s)\mathcal I(p,q,s)\bigg|_{p=q=s=1},
    \label{eq:c-afromI}
\end{equation}
where $\lambda$ is an undetermined constant.  This ambiguity arises because the combination $I_{2,2}-I_2$ vanishes for the singlet and (anti-)triplet representations.

\begin{table}[t]
\begin{center}
  \begin{tabular}{| l | l | l | l |}
    \hline
    \ Multiplet & \ Shortening Condition\ & \ $\mathcal D(p,q,s)\mathcal{I}_R(p,q,s)$\ &\ $2^5\cdot6!\delta(c-a)$ \\
    \hline
    \hline
    \ $A[0,0,0;k]$ \ & \ $\Delta=2k+6$ \ & \ $-q^{\hat\Delta} \, \chi_{(0,0)}(s,p)$ \ &$-1$\\ \hline
    \ $B[0,0,0;k]$ \ & \ $\Delta=2k+4$ & \ $q^{\hat\Delta} \, \chi_{(0,1)}(s,p)$ \ &\ $3+90(\fft43\hat\Delta-2)$\ \\ \hline
    \ $C[0,0,0;k]$ \ & \ $ \Delta=2k+2$ & \ $- q^{\hat\Delta} \, \chi_{(1, 0)}(s,p)$ \ &\ $-3+90(\fft43\hat\Delta-2)$\ \\ \hline
    \ $D[0,0,0;k]$ \ & \ $\Delta=2k$ & \ $q^{\hat\Delta} \, \chi_{(0, 0)}(s,p)$ \ &\ $1$\ \\ \hline
  \end{tabular}
\end{center}
\caption{The single-trace index and holographic $\delta(c-a)$ for maximum spin-two $(1,0)$ short multiplets.  The $\delta(c-a)$ results are taken from \cite{Liu:2017ruz}, but are given here in terms of $\hat\Delta\equiv\Delta-\fft12k$.
\label{tb:c-aindex}}
\end{table}

\section{Discussion}

The six-dimensional $(1,0)$ SCFT has three independent central charges, which we have denoted $a$, $c$ and $c'$.  For large-$N$ theories admitting a holographic dual, we have demonstrated that the $\mathcal O(1)$ contributions, $\delta a$ and $\delta(c-a)$ can be obtained from the single-trace index by the action of the differential operators given in (\ref{eq:afromI}) and (\ref{eq:c-afromI}), at least up to one undetermined coefficient for $\delta(c-a)$.  In order to fix this coefficient, we would have to work out $\delta(c-a)$ for at least one higher-spin multiplet.  The most obvious choice would be $C[1,0,0;k]$, however this would require the investigation of three additional fields, transforming in the higher-spin $(3,0,0)$, $(2,1,0)$ and $(2,0,1)$ representations of $SU(4)$.

In holographic theories, the leading order behavior of the central charges scales as $N^3$, and the first subleading corrections arise at $\mathcal O(N)$.  So in practice the $\mathcal O(1)$ terms that we have identified from the single-trace index are rather small corrections.  Nevertheless, their structure can provide a hint at a more complete relationship between the full index and central charges.  The full index, of course, differs from the single-trace index, but can be related through the plethystic exponential.  As in the AdS$_5$/CFT$_4$ case considered previously \cite{Ardehali:2014zba,Ardehali:2014esa}, we expect that the connection of $\delta a$ and $\delta(c-a)$ to the single-trace index generalizes in terms of the high-temperature structure of the full index \cite{DiPietro:2014bca,Ardehali:2015hya,Assel:2015nca,Ardehali:2015bla,DiPietro:2016ond}.

What we mean here by the high-temperature limit comes from the connection between the superconformal index and the supersymmetric partition function on $S^n\times S^1$ \cite{Lorenzen:2014pna,Assel:2015nca}:
\begin{equation}
    \mathcal I(\beta)=e^{\beta E_{\mathrm{susy}}}Z_{S^n\times S^1_\beta},
\end{equation}
where $E_{\mathrm{susy}}$ is the supersymmetric Casimir energy and the inverse temperature $\beta$ is associated with the radius of $S^1$.  As highlighted in \cite{DiPietro:2014bca,Assel:2014paa,Ardehali:2015hya,Ardehali:2015bla}, the four-dimensional index has a high-temperature expansion of the form
\begin{equation}
    \log\mathcal I(\beta)\sim\fft{16\pi^2(c-a)_{\mathrm{shifted}}}{3\beta}+\mathrm{dim}\,\mathfrak h_{qu}\log\left(\fft{2\pi}\beta\right)+\beta E+\cdots,
    \label{eq:hiT}
\end{equation}
where $(c-a)_{\mathrm{shifted}}$ is related to a possible displacement of the minimum of the effective potential away from the origin \cite{Ardehali:2015bla,DiPietro:2016ond}.  For semi-simple gauge theories where the effective potential does not have any flat directions, the coefficient $E$ of the linear term in $\beta$ is the four-dimensional supersymmetric Casimir energy, $E_{\mathrm{susy}}=\fft4{27}(3c+a)$ \cite{Kim:2012ava,Assel:2014paa}.  It is this term that is connected to the holographic one-loop computation of $\delta a$ and $\delta c$ \cite{Ardehali:2015hya} when generalized to the squashed sphere.

In six dimensions, the high-temperature expansion of the index instead takes the form \cite{DiPietro:2014bca}
\begin{equation}
    \log\mathcal I(\beta)\sim\fft{8\pi^4}{9\beta^3}C_0+\fft{\pi^2}{6\beta}C_1+\cdots+\beta E_{\mathrm{susy}}+\ldots,
\end{equation}
where it was suggested that the factors $C_0$ and $C_1$ are related to the 't~Hooft anomaly coefficients
\begin{equation}
    \mathcal I_8=\fft1{4!}[\alpha c_2(R)^2+\beta c_2(R)p_1(T)+\gamma p_1(T)^2+\delta p_2(T)],
\end{equation}
by
\begin{equation}
    C_0=\gamma+\ft14\delta,\qquad C_1=\ft92\beta-8\gamma+\delta.
\end{equation}
While the holographic $\delta a$ and $\delta(c-a)$ are related to $E_{\mathrm{susy}}$, and therefore do not constrain $C_0$ and $C_1$, one may hope that aspects of the holographic dual can nevertheless refine our understanding of these terms.  In any case, we note that, while $C_0$ receives non-vanishing contributions from free $(1,0)$ scalar and tensor multiplets \cite{Benvenuti:2016dcs,Bak:2016vpi,Gustavsson:2018sgi}, it nevertheless vanishes in the $(2,0)$ theory \cite{Bhattacharya:2008zy,Kim:2012qf,Kim:2013nva,DiPietro:2014bca,Kim:2016usy}.  This leaves us with the question of whether any additional meaning can be attributed to $C_0$.  One way to distinguish $(1,0)$ from $(2,0)$ theories is the vanishing of the $c'$ central charge in the latter.  However, the relation \cite{Cordova:2015fha,Beccaria:2015ypa,Yankielowicz:2017xkf,Beccaria:2017dmw}
\begin{equation}
    a=-\fft1{72}(\alpha-\beta+\gamma+\ft38\delta),\qquad c-a=-\fft\delta{192},\qquad c'=\fft1{432}(\beta-2\gamma+\ft12\delta),
\end{equation}
demonstrates that this cannot be the complete story.  Likewise, the relation between $C_1$ and the central charges is not clear either. Finally, it has been conjectured in \cite{bobev:2015kza} that the supersymmetric Casimir energy can be related to the anomaly polynomial, $\mathcal I_8$, and it would be interesting to see if this could provide additional input on the structure of the high-temperature expansion of the index. These issues merit further study, as their resolution will lead to a deeper understanding of six-dimensional SCFTs.

\acknowledgments

We wish to thank A.~Arabi Ardehali, F.~Larsen and P.~Szepietowski for stimulating discussions and N.~Bobev for interesting comments. We especially wish to thank A. Arabi Ardehali for clarifying the nature of the high-temperature expansion of the index given in (\ref{eq:hiT}).
This work was supported in part by the US Department of Energy under Grant No.~DE-SC0007859.

\bibliography{cite.bib}
\end{document}